*Title:* **Benchmarking Ten Codes Against the Recent GSI Measurements of the Nuclide Yields from $^{208}$Pb, $^{197}$Au, and $^{238}$U + p Reactions at 1 GeV/nucleon**

*Author(s):* Stepan G. MASHNIK, Richard E. PRAEL, Arnold J. SIERK, Vyacheslav F. BATYAEV, Svetlana V. KVASOVA, Ruslan D. MULAMBETOV, and Yury E. TITARENKO

*Submitted to:* International Conference on Nuclear Data for Science and Technology, October 7-12, 2001, Tsukuba, Japan

http://lib-www.lanl.gov/la-pubs/00818527.pdf

# Benchmarking Ten Codes Against the Recent GSI Measurements of the Nuclide Yields from $^{208}$Pb, $^{197}$Au, and $^{238}$U + p Reactions at 1 GeV/nucleon

Stepan G. MASHNIK[1,*], Richard E. PRAEL[1], Arnold J. SIERK[1],
Vyacheslav F. BATYAEV[2], Svetlana V. KVASOVA[2], Ruslan D. MULAMBETOV[2], and Yury E. TITARENKO[2],

[1] *Los Alamos National Laboratory, Los Alamos, NM 87545, USA*
[2] *Institute for Theoretical and Experimental Physics, B.Cheremushkinskaya 25, 117259 Moscow, Russia*

A qualitative and quantitative comparison of the recent GSI measurements of the nuclide yields from $^{208}$Pb and $^{238}$U at 1 GeV/nucleon and $^{197}$Au at 800 MeV/nucleon interactions with protons with the codes LAHET (with both ISABEL and Bertini options), CEM95, CEM97, CEM2k, CASCADE, CASCADE/INPE, YIELDX, HETC, and INUCL is presented. The predictive power of these codes is reasonable for nuclides in the near spallation region but is worse for deep spallation and much worse in the fission region. None of these codes agree well with the data in the whole mass region of product nuclides and all must be improved to become reliable tools for accelerator-driven applications.

## I. Introduction

Current designs of hybrid reactor systems driven with high current accelerators require information about residual nuclides that are produced by high-energy protons interacting in the target and in structural materials. It is both physically and economically impossible to measure all necessary data, which is why reliable models and codes are needed.

The recent GSI measurements performed using inverse kinematics for interactions of $^{208}$Pb[1,2] and $^{238}$U[3] at 1 GeV/nucleon and $^{197}$Au at 800 MeV/nucleon[4] with liquid $^{1}$H provide a very rich set of cross sections for production of practically all possible isotopes from these reactions in a "pure" form, *i.e.*, individual cross sections from a specific given bombarding isotope (or target isotope, when considering reactions in the usual kinematics, p + A). Such cross sections are much easier to compare to models than the "camouflaged" data from $\gamma$-spectrometry measurements. These are often obtained only for a natural composition of isotopes in a target and are mainly for cumulative production, where measured cross sections contain contributions nor only from a direct production of a given isotope, but also from all its decay-chain precursors.

In the present work, we perform a benchmark of ten different codes widely used in current applications, namely LAHET (with both ISABEL and Bertini options),[5] CEM95,[6] CEM97,[7] CEM2k,[8] CASCADE,[9] CASCADE/INPE,[10] YIELDX,[11] HETC,[12] and INUCL[13] against the new GSI data to evaluate the predictive power of these codes and to consider ways to further improve them.

## II. Results and Discussions

To evaluate the predictive power of the tested codes, we calculated with the default options and without any changes or modifications for all codes except CEM2k, which is an improved version of its precursor, CEM97, and was developed to describe as well as possible these GSI data and other similar measurements (see details in[8,14]).

The limited size of the present paper does not allow us to describe all results of our benchmark, therefore we show here only several typical examples. **Figure 1** shows a comparison of the GSI data for p(1 GeV) + $^{208}$Pb with predictions by the LAHET code with both ISABEL and Bertini options. **Figure 2** shows similar comparisons with results from CEM97 and HETC, while **Fig. 3** shows sample of results from CEM2k. Similar figures for p(1 GeV) + $^{208}$Pb with results from INUCL, CASCADE, CASCADE/INPE, and YIELDX may be found in[15] and from CEM95, in.[8] Figures with comparisons of results from several codes tested here for the reactions p(800 MeV) + $^{197}$Au and p(1 GeV) + $^{238}$U may be found in[8,14] and will be shown in our poster at this conference. Tables with quantitative information concerning the agreement of the tested codes with experimental data may be found in[16,17] and will be shown in our presentation at this conference.

The ability of codes to simulate fission processes is an important criterion for their application when the target is heavy enough to fission. Among the codes benchmarked here, LAHET, CASCADE, INUCL, CASCADE/INPE, and YIELDX simulate both spallation and fission products. The CEM95, CEM97, CEM2k, and HETC codes simulate spallation only and do not calculate the process of fission, and do not provide fission fragments and a further possible evaporation of particles from them. When, during a Monte-Carlo simulation of the compound stage of a reaction these codes encounter a fission, they simply remember this event (that permits them to calculate the fission cross section and fissility) and finish the calculation of this event without a subsequent calculation of fission fragments. Therefore, results from HETC, CEM95, CEM97, and CEM2k reflect the contribution to the total yields of the nuclides only from deep spallation processes of successive emission of particles from the target, but do not contain fission products. To be able to describe nuclide production in the fission region, these codes have to be extended by incorporating a model of high energy fission (e.g., in the transport code MCNPX,[18] where CEM97 and CEM2k are used, they are supplemented by the RAL fission model[19]).

From the results presented here and in our papers cited above, one can see that though practically all the tested codes describe quite well the production of several heavy isotopes near the targets (we calculate p + $^{208}$Pb, $^{197}$Au, and $^{238}$U; therefore, for us $^{208}$Pb, $^{197}$Au, and $^{238}$U are targets not projectiles as in the GSI measurements), except for CEM2k they do not reproduce correctly the cross sections for lighter isotopes in the deep spallation region. The disagreement increases with increasing distance from the target, and all calculated curves are shifted to the heavy mass direction, just as was obtained by the authors of the GSI measurements with all the codes they used. In other words, this means that for a given final isotope (Z), these codes predict emission of too few neutrons. Most of the neutrons are emitted at the evaporation stage of a reaction. Accordingly, one way to increase the number of emitted neutrons would be to increase the evaporative part of a reaction (relative to preequilibrium). This (and other improvements and refinements; see details in[8, 14]) we did when developing the code CEM2k which agrees best of these tested codes with the GSI (and ITEP) data in the spallation region. However, CEM2k is inapplicable in the fission-product region, as to date it has no model of fission-fragment formation.

---

* Corresponding author, Tel. +1-505-667-9946, FAX +1- 505-667 -1931, E-mail: mashnik@t2y.lanl.gov

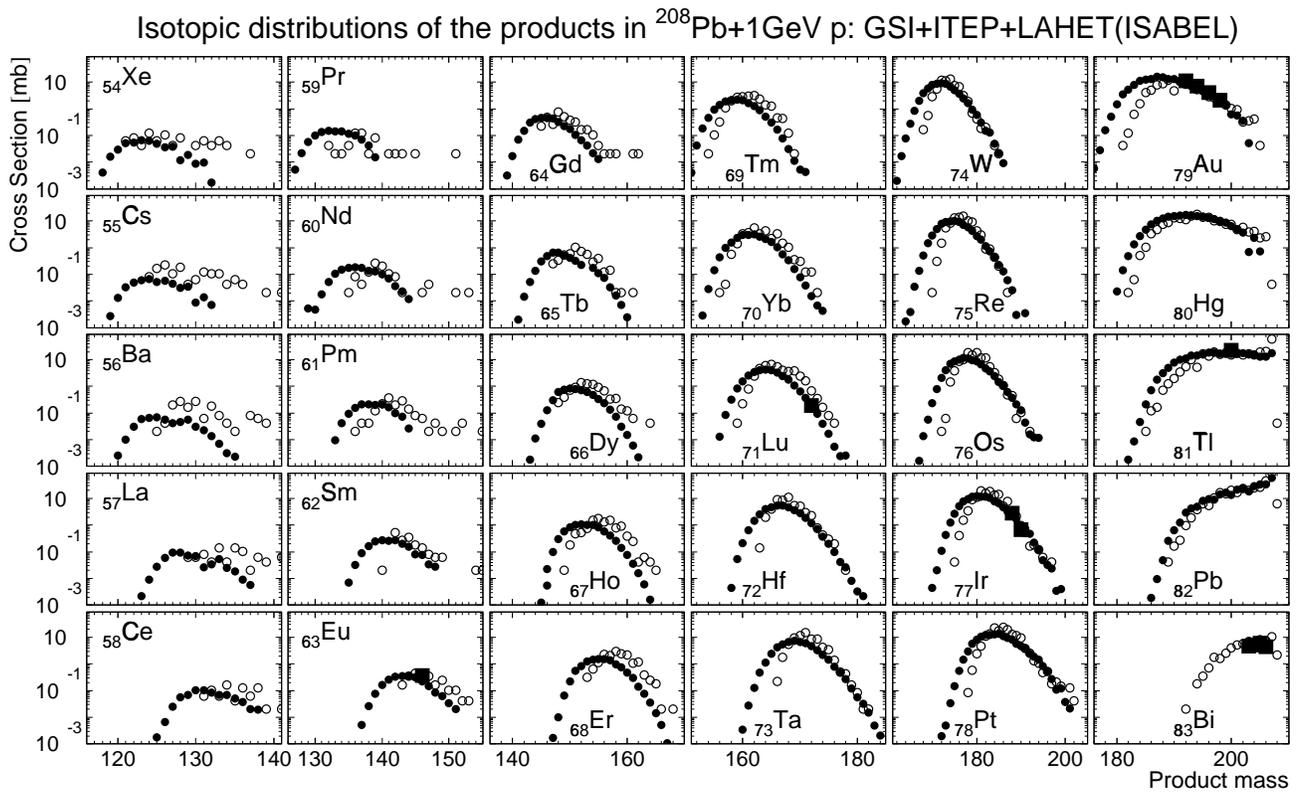

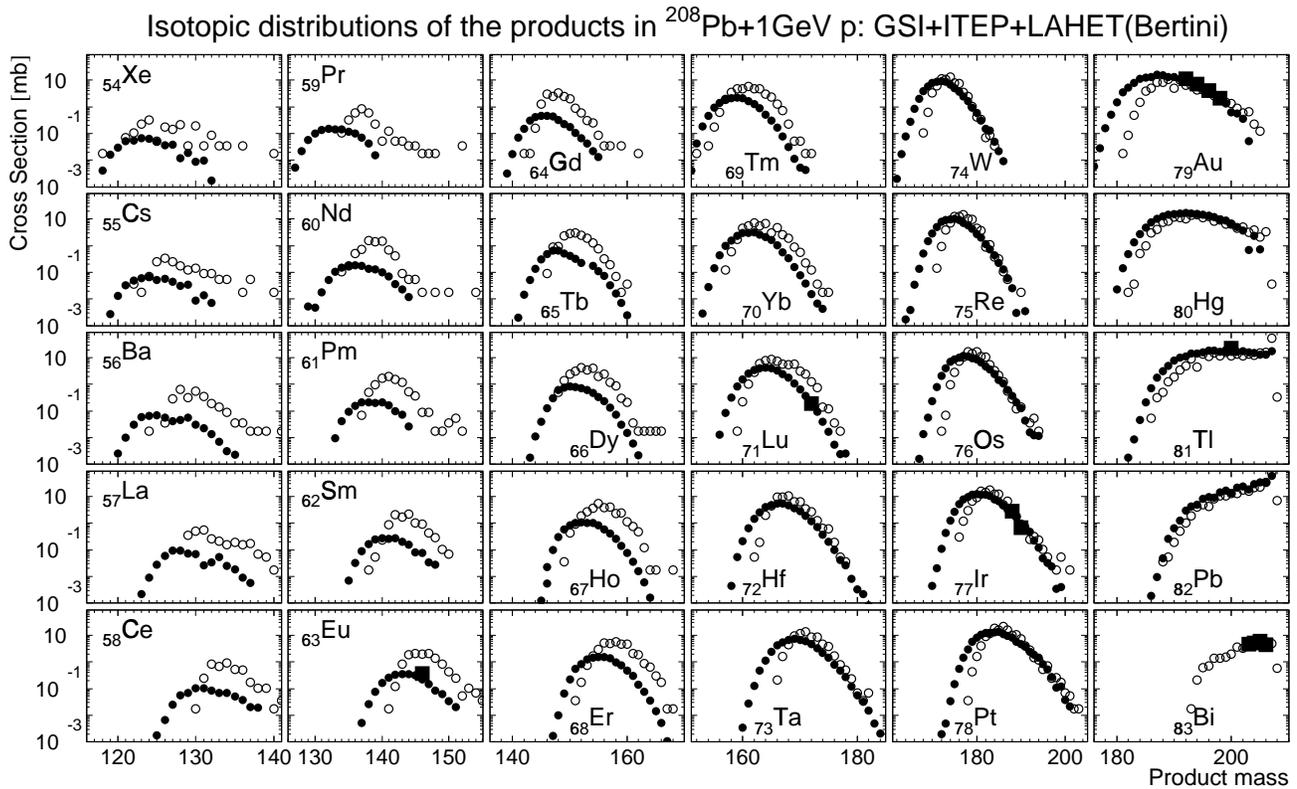

**Fig. 1** Comparison between the GSI data on nuclide yields from p (1 GeV) + $^{208}$Pb (filled circles) and calculations by LAHET-ISABEL and LAHET-Bertini (open circles); filled squares show experimental data for the same reactions measured by the $\gamma$-spectrometry method at ITEP by Titarenko *et al.*[16])

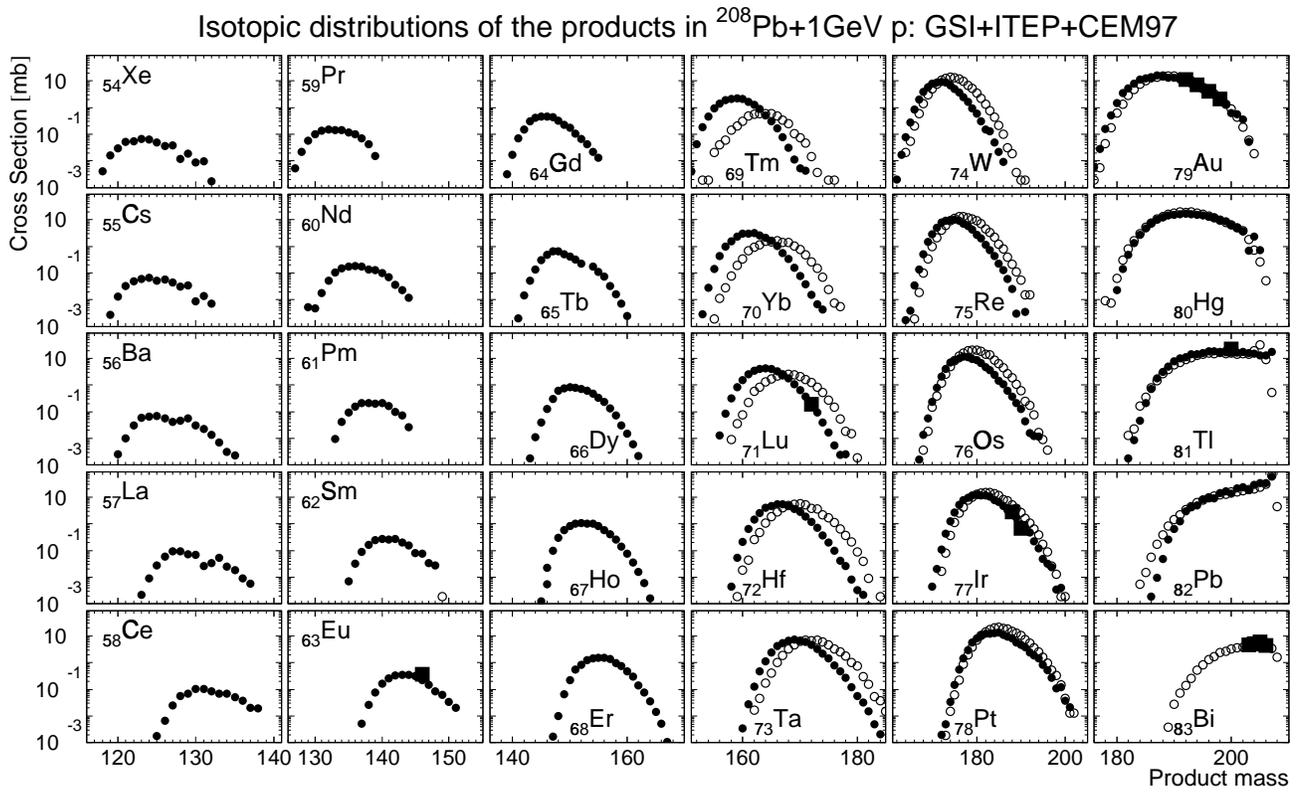
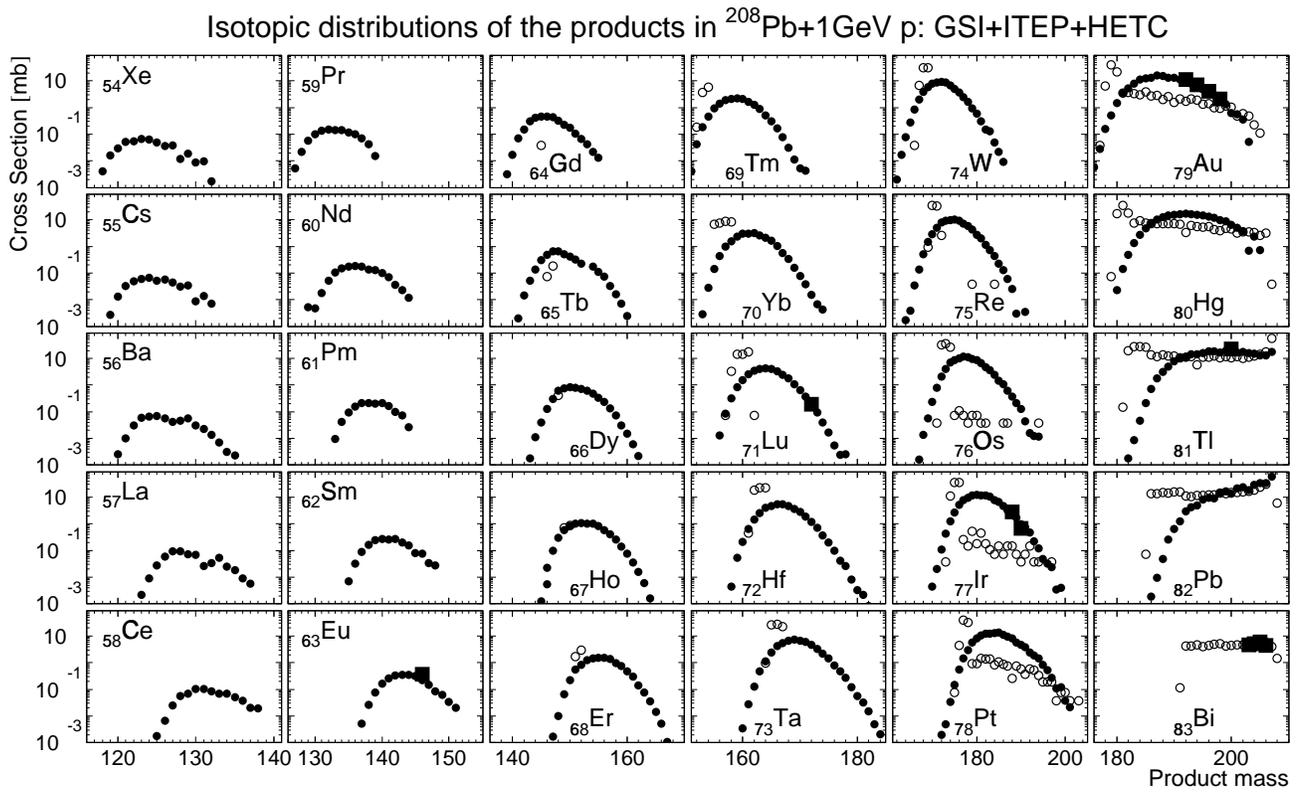

**Fig. 2** The same as in Fig. 1 but compared with CEM97 and HETC calculations.

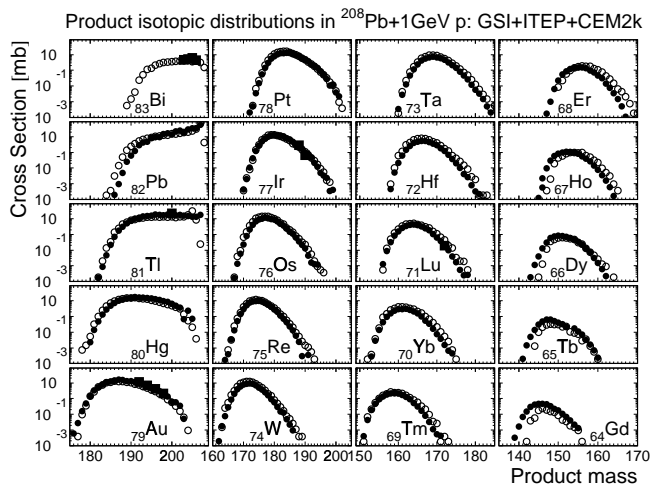

**Fig. 3** The same as in Fig. 1 but compared with CEM2k calculation.

## III. Conclusion

The predictive power of the tested codes is different but was found to be reasonable for most of the nuclides in the near spallation region (for nuclides not too far from targets), though none of the benchmarked codes agree well with the data in the whole mass region of product nuclides and all should be improved further. On the whole, the predictive power of all benchmarked codes for the data in the fission product region is worse than in the spallation region; therefore we conclude that development of better models for fission-fragment formation is of first priority.

**Acknowledgment**

We thank Drs. T. Enqvist and B. Mustapha for providing us with the cross sections measured at GSI. The work was supported by the U. S. Department of Energy and by the ISTC Project #839B.